\documentclass[conference]{IEEEtran}
\IEEEoverridecommandlockouts
\usepackage{cite}
\usepackage{amsmath,amssymb,amsfonts}
\usepackage{algorithmic}
\usepackage{graphicx}
\usepackage{tabulary}
\usepackage{textcomp}
\usepackage{xcolor}
\usepackage{fix-cm}
\usepackage{subfigure} 
\usepackage{tikz}
\usepackage{pgfplots}
\usepackage{booktabs}
\usepackage[hidelinks]{hyperref}
\usetikzlibrary{fit}
\def\BibTeX{{\rm B\kern-.05em{\sc i\kern-.025em b}\kern-.08em
    T\kern-.1667em\lower.7ex\hbox{E}\kern-.125emX}}

\newcommand{\linebreakand}{%
  \end{@IEEEauthorhalign}
  \hfill\mbox{}\par
  \mbox{}\hfill\begin{@IEEEauthorhalign}
}

\newcommand\copyrighttext{%
	\footnotesize \textcopyright 2020 IEEE. Personal use of this material is permitted.
	Permission from IEEE must be obtained for all other uses, in any current or future
	media, including reprinting/republishing this material for advertising or promotional
	purposes, creating new collective works, for resale or redistribution to servers or
	lists, or reuse of any copyrighted component of this work in other works.
	DOI: \href{https://doi.org/10.1109/MetroInd4.0IoT48571.2020.9138205}{10.1109/MetroInd4.0IoT48571.2020.9138205}}
\newcommand\copyrightnotice{%
	\begin{tikzpicture}[remember picture,overlay]
		\node[anchor=south,yshift=10pt] at (current page.south) {\fbox{\parbox{\dimexpr\textwidth-\fboxsep-\fboxrule\relax}{\copyrighttext}}};
	\end{tikzpicture}%
}

\begin{document}

\title{Quality Assurance of Weld Seams Using Laser Triangulation Imaging and Deep Neural Networks}

\author{\IEEEauthorblockN{Andreas Spruck}
\IEEEauthorblockA{\textit{Multimedia Communications and} \\ \textit{Signal Processing} \\
\textit{University of Erlangen-N\"urnberg}\\
Erlangen, Germany \\
andreas.spruck@fau.de}
\and
\IEEEauthorblockN{J\"urgen Seiler}
\IEEEauthorblockA{\textit{Multimedia Communications and} \\ \textit{Signal Processing} \\
\textit{University of Erlangen-N\"urnberg}\\
Erlangen, Germany \\
juergen.seiler@fau.de}
\and
\IEEEauthorblockN{Michael Roll}
\IEEEauthorblockA{\textit{Autotech Engineering Deutschland GmbH} \\
Bielefeld, Germany \\
michael.roll@de.gestamp.com}
\and
\IEEEauthorblockN{Thomas Dudziak}
\IEEEauthorblockA{\textit{Autotech Engineering Deutschland GmbH} \\
Bielefeld, Germany \\
thomas.dudziak@de.gestamp.com}
\and
\IEEEauthorblockN{Jürgen Eckstein}
\IEEEauthorblockA{\textit{Autotech Engineering Deutschland GmbH} \\
Bielefeld, Germany \\
juergen.eckstein@de.gestamp.com}
\and
\IEEEauthorblockN{Andr\'e Kaup}
\IEEEauthorblockA{\textit{Multimedia Communications and} \\ \textit{Signal Processing} \\
\textit{University of Erlangen-N\"urnberg}\\
Erlangen, Germany \\
andre.kaup@fau.de}
}

\maketitle
\copyrightnotice

\begin{abstract}
In this paper, a novel optical inspection system is presented that is directly suitable for Industry 4.0 and the implementation on IoT-devices controlling the manufacturing process. The proposed system is capable of distinguishing between erroneous and faultless weld seams, without explicitly defining measurement criteria . The developed system uses a deep neural network based classifier for the class prediction. A weld seam dataset was acquired and labelled by an expert committee. Thereby, the visual impression and assessment of the experts is learnt accurately. In the scope of this paper laser triangulation images are used. Due to their special characteristics, the images must be pre-processed to enable the use of a deep neural network. Furthermore, two different approaches are investigated to enable an inspection of differently sized weld seams. Both approaches yield very high classification accuracies of up to 96.88\%, which is competitive to current state of the art optical inspection systems. Moreover, the proposed system enables a higher flexibility and an increased robustness towards systematic errors and environmental conditions due to its ability to generalize. A further benefit of the proposed system is the fast decision process enabling the usage directly within the manufacturing line. Furthermore, standard hardware is used throughout the whole presented work, keeping the roll-out costs for the proposed system as low as possible.
\end{abstract}


\section{Introduction}
Due to recent advances in the field of production engineering, it is nowadays possible to produce components precise enough such that they approach the fail limit very accurate. 
Nevertheless, as a very high quality should be guaranteed to the customer, it is essential to inspect the produced items before shipping. As quality inspection performed by humans is quite expensive and tiring for the workers, numerous automated in-line inspection systems were developed within the last years \cite{Pasinetti2018, Massaro2019}. Moreover, such in-line systems bear great potential under the aspect of Industry 4.0 and IoT, as the outcome of the inspection can be directly fed back to the preceding manufacturing process. Thereby, process parameters can be steered with a very low delay, decreasing the number of rejections. Nevertheless, all of these systems have in common that they check strictly defined parameter patterns and classify the item based on the outcome of those measurements as faultless or erroneous. Aberrations of other parameters will not lead to a change of the decision, whereas side effects, that corrupt the measurement, will have a negative effect on the decision process, which can lead to misclassification \cite{SCHWENKE}. Such effects can be a displacement of the item within the measurement unit or a batch-change of the material, causing a shift of acquisition characteristics, which results in a corruption of the measured parameters, without any physical change of the item. A very challenging task regarding such interferences is the assessment of weld seams \cite{Pasinetti2018}. The difference between faultless and erroneous seams is often hard to distinguish on image data.
As weld seams often build a crucial part of larger components, a failure of a weld seam often causes drastic malfunctions. Thus, the inspection of weld seams is in many cases inevitable. 
The objective is to overcome the problems caused by interferences during the image acquisition by using deep neural networks (DNN). 
Neural networks are able to generate a more holistic model of the item, making it robust towards such errors. A further advantage of the proposed DNN-based system is that the explicit parametrization of the measurement task is omitted and replaced by an expert vote on an initial training dataset.\\
Section~\ref{sec:AOI} gives a short overview of automatic inspection systems for weld seams. Section~\ref{sec:DNN-insp} briefly describes the general principle of the proposed quality assurance system based on deep neural networks. Section~\ref{sec:Database} introduces the database used throughout this paper. In Section~\ref{sec:Eval}, the results obtained with this system and the different training strategies are presented. Section~\ref{sec:Conclusion} concludes the paper.

\section{Automated Inspection of Weld Seams}
\label{sec:AOI}

Nowadays many different automated optical inspection (AOI) systems exist. Depending on the field of application, their technical realization can vary drastically. The acquisition system is determined according to the requirements, regarding measurement precision and the material of the item that is to be inspected. In many cases a camera or X-ray apparatus, but also IR cameras or ultrasonic sensors are used for data acquisition \cite{Stavridis2018}. For more sophisticated applications, a laser triangulation sensor might be used, which enables far more precise measurements due to a finer resolution \cite{Chu2017, Minnetti2019}, even a combination of various measurement techniques can be applied \cite{zaiss2017new}. The geometry of the item of interest is an important factor for the implementation of the acquisition system. For flat surfaces or items without any shielded or shadowed areas, the acquisition system can be mounted at a fixed position. 
For more complex shaped components either the acquisition system is mounted to a flexible platform or the items have to be moved in front of the acquisition system in such a way that the inspection of every potentially critical point can be accomplished. Therefore, the acquisition system might be mounted to an industrial robot or the items are moved by such a robot in front of the measurement unit. Again a combination of both approaches is possible \cite{pernkopf2003image,Massaro2019,Pasinetti2018}.\\
Even though the acquisition system and process might vary strongly, the general principle of the inspection and classification is similar for all current systems. Based on the recorded data, certain predefined measurements are performed and specified object dimensions are extracted. Depending on the outcome of these measurements, a decision is made by comparing the results to a predefined pattern and checking whether they lie within a fixed tolerance range. The predefined pattern is specified with respect to an ideal weld seam and the allowed tolerance range. Based on the examinations, an overall decision is made 
\cite{shirvaikar2006trends}. If even only one of these measurements is corrupted by any side effect during the measurement or the image acquisition, the whole process of decision making is corrupted. This makes the current state of the art inspection systems very susceptible to systematic errors and changing environmental conditions, which might result in a high number of undesired false-negatives \cite{Sun2018}.

\section{DNN-Based Inspection of Weld Seams}
\label{sec:DNN-insp}
\begin{figure}
\centering
\resizebox{0.45\textwidth}{!}{
\begin{tikzpicture}
\node[draw, rectangle] at (0,8) (LS) {Robot-mounted laser scanner};

\draw [ultra thick] (-2.25,3.75) rectangle (2.25,7) node[align=left, anchor= north west] {Data \\ pre-processing};
\node[draw, rectangle] at (0,6.5) (GC) {Gamma correction};
\node[draw, rectangle] at (0,5.75)(N) {Normalization};
\node[draw, rectangle] at (0,5) (con) {16 bit to 8 bit conversion};
\node[draw, rectangle] at (0,4.25) (res) {Resizing};

\node[draw, rectangle] at (0,3.25) (DNN) {Inception V3 network};

\node[draw, rectangle] at (0,2.5) (QA) {Quality decision};

\draw [->] (LS) -- (GC) node[pos=0.25, right] {Intensity image};
\draw [->] (GC) -- (N);
\draw [->] (N) -- (con);
\draw [->] (con) -- (res);
\draw [->] (res) -- (DNN);
\draw [->] (DNN) -- (QA);
\end{tikzpicture}
}
\caption{Flowchart of the proposed optical inspection system.}
\label{fig:flowchart}
\end{figure}
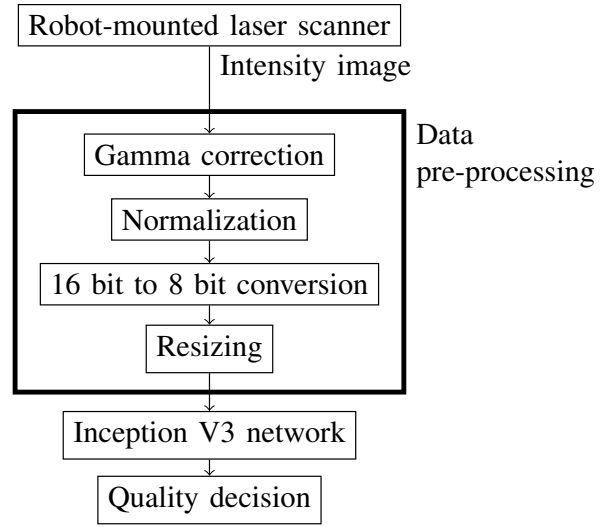

In this paper, we propose an alternative to traditional inspection systems. A flowchart of our proposed system is given in Fig.~\ref{fig:flowchart}. With the recent progress in the field of machine learning, specifically in the field of deep neural networks, powerful tools were created, whose great advantage lies in the broad variety of possible application fields. The strength of DNNs is to discover high dimensional features within image data and exploit these for the task they are trained for. 

\begin{figure}[htbp]
\centering
\subfigure[Example of a faultless weld seam.]{\label{fig:example_seams_a} \includegraphics[width=0.45\textwidth]{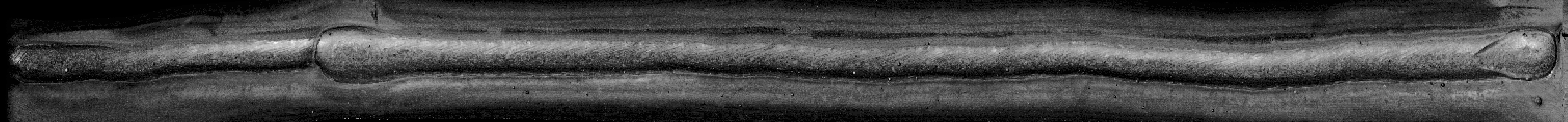}}
\\ 
\subfigure[Example of an erroneous weld seam.]{\label{fig:example_seams_b} \includegraphics[width=0.45\textwidth]{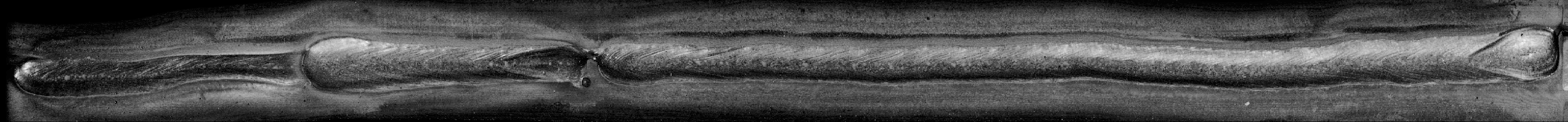}}
\caption{Exemplary plot of a faultless and an erroneous weld seam.}
\label{fig:example_seams}
\end{figure}

Our proposed system, as depicted in Fig.~\ref{fig:flowchart}, incorporates a state-of-the-art DNN to classify the acquired images of weld seams. As the images of the weld seams are scanned using a laser triangulation sensor they strongly differ from usual image data. Therefore, pre-processing of the image data is necessary. The special characteristics of the image data as well as the data pre-processing are explained in further detail in Section~\ref{sec:Database}. \\
Within the proposed system, the risk of false-classification due to systematic errors is minimized without increasing the number of false-positives by overcoming the fixed parametrized pattern of the traditional inspection systems. Interferences such as displacements during the acquisition or changes in the used material may not lead to misclassification. By including such interferences in the training dataset of the neural network, the network is made more robust against systematic errors. A further advantage of the DNN-based system is that the whole previously installed inspection equipment can be reused for the newly proposed system. Only the strictly parametrized pattern is replaced by a more sophisticated and holistic DNN-classifier, keeping the hardware of the system untouched. 
In the proposed system 
we modified an existing laser triangulation based inspection system in such a way that the intensity images acquired by this system were used as input for the newly introduced DNN-classifier. By reusing the existing infrastructure the roll-out costs are kept as low as possible. The acquisition system is implemented in such a way that the laser triangulation sensor is mounted to an industrial robot, which moves the sensor over the component along the seam. By this procedure, the resulting intensity image shows a long straight weld seam even though the original geometry of the seam is round. Fig.~\ref{fig:example_seams} shows an example of a faultless and an erroneous weld seam, as they are used here. Due to this procedure, the resulting images have a strongly unbalanced aspect ratio, as the length of the seam is depicted along the horizontal axis. The used deep neural network used for classification is an implementation of the Inception V3 network \cite{Szegedy2016}. This deep neural network offers a good compromise between computational complexity and classification accuracy for different application cases, as it is shown in \cite{Junayed2018}. 

\begin{figure}[htbp]
\centering
\subfigure[Example of a shrunk weld seam.]{\label{fig:example_resized_seams_a} \includegraphics[width=0.2\textwidth]{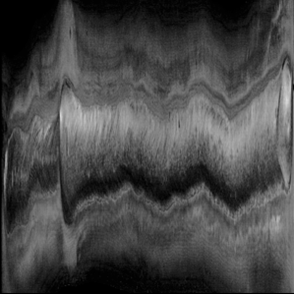}} \hspace{0.5cm} 
\subfigure[Example of scaled weld seam.]{\label{fig:example_resized_seams_b} \includegraphics[width=0.2\textwidth]{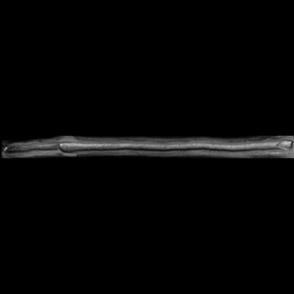}}
\caption{Comparison of a shrunk weld seam without fixed aspect ratio resulting in a strong compression along the horizontal axis and a scaled weld seam with fixed aspect ratio.}
\label{fig:example_resized_seams}
\end{figure}

A challenge during this task was that, depending on the type of seam that was to be inspected, the image size varies within the different seam types, whereas the used neural network always expects a fixed input size of the image of 299×299 pixels. Due to the constraint that the same setup should be able to handle any arbitrary seam, 
two different options are examined here. The first option is to resize the image to a size of 299×299 pixels, which typically results in a strong compression along one axis, as shown in Fig.~\ref{fig:example_resized_seams_a}. The second option is to scale the image of the seam, while keeping the aspect ratio of the original image, which is depicted in Fig.~\ref{fig:example_resized_seams_b}. During both approaches, we used a bicubic interpolation for resizing.

\section{Properties of the Acquired Weld Seam Database}
\label{sec:Database}
A crucial requirement for the DNN-based system is a representative database for training that clearly shows the possible erroneous parts of the object. Furthermore, a sufficient number of labelled training images is required, as the quality of the final classification result strongly depends on the training data. That does not only refer to the number of available examples, but also to the quality of the assigned labels. The required amount of training data strongly depends on the complexity of the inspection task. The more complex the inspection task is, the more training data is required. Nevertheless, we were able to show that reasonable results can be achieved with a manageable amount of data, as the used dataset yielded only 616 weld seam scans. The weld seams were scanned with a laser triangulation sensor. During the scan, the laser triangulation sensor is moved along the seam. Thus, the dataset only contains grayscale intensity images with strongly unbalanced aspect ratio. Even though the training process of the neural network is quite time consuming compared to the evaluation and might seem as a disadvantage at first sight, the parametrization of the inspection pattern of the traditional inspection systems is omitted, as well as the fine tuning of the parameters during the application as reaction on changing production conditions.\\
In order to receive training data with a high label quality, an expert committee consisting of five welding professionals inspected and labelled each intensity image used throughout the experiments. Apart from the training data, this committee labelled the validation and the test data, as well. To keep the label quality high for both approaches, the expert committee examined and labelled the seams at original high resolution before resizing. Labelling the images at high resolution is important, because after resizing the shape of the seam is either drastically deformed, due to the large compression along one axis or very small, such that a meaningful labelling by the expert group cannot be guaranteed for the resized images. %
To overcome this problem and keep the label quality during training, validation, and testing as high as possible, the weld seams were labelled at full resolution. 
In order to enlarge the available dataset, data augmentation methods are used \cite{Krizhevsky2012}. Each seam is reflected along its horizontal and vertical axis. Due to that, the size of the dataset could be increased by a factor of four. A challenge during the experiments for both approaches is that the class distribution is very unbalanced within the available dataset. This is due to the fact that far more error-free than erroneous components have been produced. During the partitioning into the three datasets, training, validation, and test dataset, special attention was paid to an equal distribution of the classes within the validation and test dataset. Hence, the resulting accuracy values can be easily and unambiguously interpreted, as they are not corrupted by a biased dataset. Furthermore, each scan of a seam solely occurs in one dataset, which also includes all augmented versions of it. By that, the datasets can be kept independent from each other. The whole dataset consists of 553 faultless and 63 erroneous seams, which results in 2212 error-free and 252 defective seams after enlargement of the dataset using the above mentioned data augmentation methods. Table~\ref{tab:dataset} shows the subdivision into training, validation, and test dataset. 
For both single-shot approaches - scaling and shrinking - the same subdivision into the three datasets is used. 

\begin{table}[htbp]
\caption{Partitioning of the used datasets.}
\begin{center}
\begin{tabulary}{0.45\textwidth}{|L|C|C|} \toprule
 & \textbf{\textit{Number of error-free seams}}& \textbf{\textit{Number of erroneous seams}}\\
\midrule
Training & $2084$ & $124$ \\
\midrule
\mbox{Validation} & $64$ & $64$ \\ \midrule
Test & $64$ & $64$ \\ \midrule[1.1pt]
Total & $2212$ & $252$ \\ \bottomrule 
\end{tabulary}
\label{tab:dataset}
\end{center}
\end{table}

The intensity images delivered by the laser triangulation sensor are recorded with 16-bit depth. For displaying and further usage, we have to convert the data to 8-bit. Therefore, the scans of the seams have to be pre-processed in order to label and process the images. The pixel values of the intensity images are normalized onto the maximum value within the scan. Additionally a gamma correction with a factor of 0.7 is applied. During the application, the labelling step can be omitted and the intensity image acquired by the laser triangulation system can be resized instantly and fed into the neural network classifier. Solely the above mentioned pre-processing operations as the conversion from 16-bit to 8-bit depth, normalization and gamma correction stay necessary. 

\section{Evaluation of the Inspection Results}
\label{sec:Eval}
During the evaluation, each model was initialized randomly and trained three times independently from each other. The results verify the reproducibility and show that very high classification accuracies can be achieved independent from the initialisation. The consensus vote of the expert committee serves as ground-truth for the evaluation of the classification performance of the deep neural network classifier. This procedure is the same for both, above-mentioned approaches, in order to guarantee a good comparability. For the training of the deep neural network, a midrange computer was used, equipped with an Intel Xeon E3-1245 v5 CPU, 32 GB RAM and a NVIDIA GeForce GTX1070 GPU with 8 GB memory. This setup was selected to further pursue the approach of building an easy and cheap retrofit system, such that smaller companies are enabled to use modern machine learning techniques without high financial risk. 

\subsection{Inspection of Shrunk Scans}
\label{subsec:eval_shrunk}
\begin{figure}[h!]
\centerline{\includegraphics[width=0.48\textwidth]{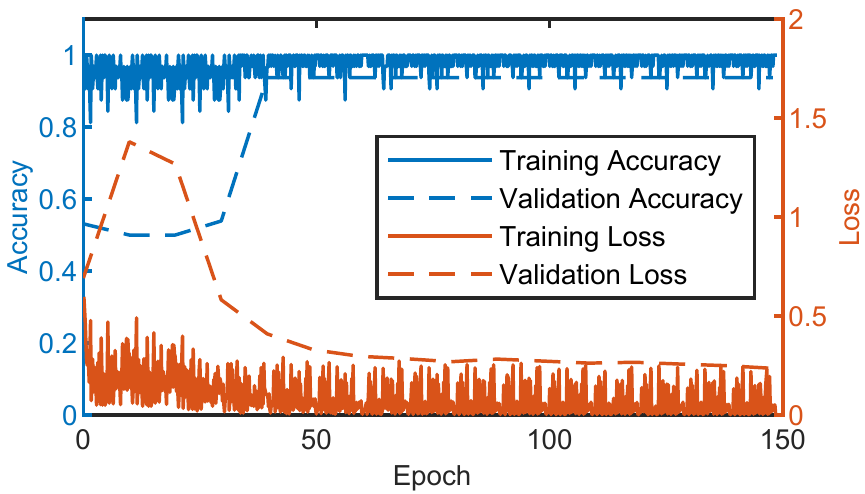}}
\caption{Training progress of the neural network using shrunk scans of the weld seams.}
\label{fig:training_shrunk}
\end{figure}

The network is trained for 150 epochs with a fixed learning rate of $1\cdot 10^{-6}$ on the datasets given in Table~\ref{tab:dataset}. 
Fig.~\ref{fig:training_shrunk} shows the training progress of the network for the shrunk scans of the seams for one exemplary run, where the desired shape of the training curves is visible, as the training and validation accuracy rise with increasing number of training steps, whereas the training and validation loss drop.

\begin{table}[htbp]
\tymin=1cm
\caption{Classification accuracy, true positive rate (TPR), true negative rate (TNR) and positive predictive value (PPV) of the neural network using shrunk scans of the weld seams.}
\begin{center}
\begin{tabulary}{0.48\textwidth}{|L|C|C|C|C|C|}
\toprule
 & \textbf{\textit{Max. validation accuracy}}& \textbf{\textit{Test accuracy}} & \textbf{\textit{ TPR }} & \textbf{\textit{ TNR }} & \textbf{\textit{ PPV }}\\
\midrule
Run 1 & $93.75\%$ & $95.31\%$ & $100\%$ & $90.63\%$ & $91.43\%$ \\ \midrule
Run 2 & $93.75\%$ & $96.09\%$ & $100\%$ & $92.19\%$ & $92.75\%$ \\ \midrule
Run 3 & $93.75\%$ & $93.75\%$ & $100\%$ & $87.50\%$ & $88.89\%$ \\ \midrule[1.1pt]
Average & $93.75\%$ & $95.05\%$ & $100\%$ & $90.11\%$ & $91.02\%$ \\ \bottomrule
\end{tabulary}
\label{tab:results_shrunk}
\end{center}
\end{table}

During the test runs, very high accuracies could be achieved. For the shrunk scans of the seams, a test accuracy of up to 96.09\% could be reached on the independent test dataset. The detailed results of all three runs are shown in Table~\ref{tab:results_shrunk}. Here it can be observed, that even for the worst of all three trained networks the test accuracy is clearly above 93\%. Considering the true positive rate (TPR) also known as recall, the true negative rate (TNR) and the positive predictive value (PPV) also known as precision it becomes visible, that no false negatives occurred during the tests while the number of false positives could be kept quite low as well. Therefore, it can be stated that the proposed solution yields in general good results and is capable of inspecting weld seams.

\subsection{Inspection of Scaled Scans}
\label{subsec:eval_scaled}
\begin{figure}[htbp]
\centerline{\includegraphics[width=0.48\textwidth]{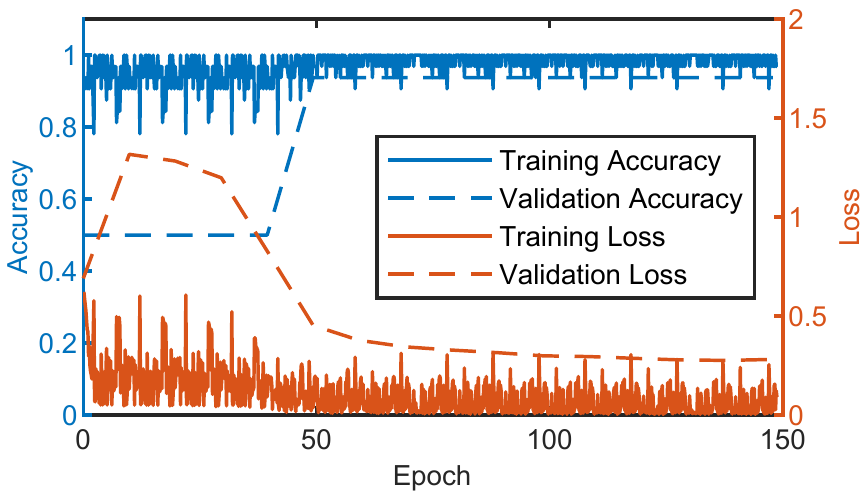}}
\caption{Training progress of the neural network using scaled scans of the weld seams.}
\label{fig:training_scaled}
\end{figure}

The training procedure for the neural networks inspecting the shrunk and the scaled scans is the same as in Section~\ref{subsec:eval_shrunk}. The network is trained for 150 epochs with a fixed learning rate of $1 \cdot 10^{-6}$. In Fig.~\ref{fig:training_scaled}, the accuracy and the loss are plotted for the training as well as for the validation over the number of training epochs. \\
During the test procedure, we could achieve even better results than with the approach using shrunk versions of the scans from Section~\ref{subsec:eval_shrunk} yielding a classification accuracy of up to 96.88\% during the final tests. As can be seen in Table~\ref{tab:results_scaled} the overall classification accuracy is almost constantly high for all three training and test runs.  Furthermore the TPR could be kept at 100\%, so there were again no false negatives with this approach. In terms of TNR and PPV the results could even be further improved as the number of false positives could be further decreased. 

\begin{table}[htbp]
\tymin=1cm
\caption{Classification accuracy, true positive rate (TPR), true negative rate (TNR) and positive predictive value (PPV) of the neural network using scaled scans of the weld seams.}
\begin{center}
\begin{tabulary}{0.48\textwidth}{|L|C|C|C|C|C|}
\toprule
 & \textbf{\textit{Max. validation accuracy}}& \textbf{\textit{Test accuracy}}& \textbf{\textit{TPR}} & \textbf{\textit{TNR}} & \textbf{\textit{PPV}}\\
\midrule
Run 1 & $93.75\%$ & $96.88\%$ & $100\%$ & $93.75\%$ & $94.12\%$ \\ \midrule
Run 2 & $93.75\%$ & $96.09\%$ & $100\%$ & $92.19\%$ & $92.75\%$ \\ \midrule
Run 3 & $93.75\%$ & $96.88\%$ & $100\%$ & $93.75\%$ & $94.12\%$ \\ \midrule[1.1pt]
Average & $93.75\%$ & $96.62\%$ & $100\%$ & $93.23\%$ & $93.66\%$ \\ \bottomrule
\end{tabulary}
\label{tab:results_scaled}
\end{center}
\end{table}

\subsection{Comparison of both Approaches}
\label{subsec:eval_comp}
As shown in Section~\ref{subsec:eval_shrunk} and \ref{subsec:eval_scaled} both approaches for single-shot weld seam inspection produce very good results. Nevertheless, the approach of using scaled versions of the weld seam scan delivers slightly better and more constant results over all three training runs. In Table~\ref{tab:results_comparison} an overview of the best achieved results on the test dataset is given for both proposed approaches as well as for the reference system.  Comparing the proposed deep neural network based inspection system to the existing automated optical inspection system it can be observed that the AOI system reaches a classification accuracy of 96.88\% on the same test dataset as the neural networks. 

\begin{table}[htbp]
\tymin=1.1cm
\caption{Comparison of the classification accuracy, true positive rate (TPR), true negative rate (TNR) and positive predictive value (PPV) on the independent test dataset with respect to expert vote.}
\begin{center}
\begin{tabulary}{0.48\textwidth}{|L|C|C|C|}
\toprule
& \textbf{\textit{Proposed system – shrunk scans}}& \textbf{\textit{Proposed system – scaled scans}}& \textbf{\textit{State of the Art reference system}}\\
\midrule
Accuracy & $96.09\%$ & $96.88\%$ & $96.88\%$ \\ \midrule
TPR & $100\%$ & $100\%$ & $100\%$ \\ \midrule
TNR & $92.19\%$ & $93.75\%$ & $93.75\%$ \\ \midrule
PPV & $92.75\%$ & $94.12\%$ & $94.12\%$ \\ \bottomrule
\end{tabulary}
\label{tab:results_comparison}
\end{center}
\end{table}

Table~\ref{tab:results_comparison} states that the proposed systems are fully competitive to the current state of the art inspection system, while enabling a clearly higher flexibility and drastically eased initialization, as the explicit parametrization of the item is omitted. By further investigation of the results, it becomes visible that both, the neural network based system as well as the AOI system misclassify the same weld seams. \\
These good classification results teamed up with the fast decision process as well as a moderate training time of 1.75 hours make the presented approach very attractive for the use during the manufacturing process, where the two main constraints are a fast and a reliable classification of the items. 
The classification of a single scan is very fast with 0.012 seconds per seam, once the model is initially loaded which takes around eight seconds and the scan being already pre-processed using the methods mentioned in Section~\ref{sec:Database}. 
Using such a DNN-based classifier it is possible to establish a certain quality standard that is directly linked to the assessment of a certain expert committee. By that the desired quality standard can be defined more widespread and independently from the place of production, as the evaluation and knowledge of the expert committee is adopted by the DNN-classifier, that imitates the committee.  

\section{Conclusion}
\label{sec:Conclusion}
This paper shows that the proposed DNN-based optical inspection system for weld seams bears a great potential for the inspection of weld seams. It is capable of producing competitive results compared to currently existing and perfectly calibrated AOI systems, while easing the initialization process. The proposed system enables the reproduction of the visual impression and assessment of the expert committee. Thus, the initialization of the proposed system is more comfortable than for current AOI, as the DNN-based quality assurance system is trained on visual examples. Thereby, the proposed system bears high flexibility due to its re-trainable decision process and bears an increased robustness towards systematic errors as well as environmental conditions. Furthermore, the eventually already installed inspection system can be reused with the proposed method, as the presented changes in the system are restricted to the classification process. 

\bibliographystyle{IEEEtran}
\bibliography{Literatur_short}

\end{document}